\documentclass[aps,twocolumn,superscriptaddress,prb]{revtex4-2}

\usepackage{bm,graphicx,textcomp,amssymb,amsmath,dcolumn,color}

\newcommand{\ket}[1]{\left|#1\right>}
\newcommand{\bra}[1]{\left<#1\right|}

\newcommand{\nn}{\nonumber\\}

\newcommand{\f}[1]{\mbox{\boldmath$#1$}}
\newcommand{\fk}[1]{\mbox{\boldmath$\scriptstyle#1$}}

\newcommand{\bea}{\begin{eqnarray}}
\newcommand{\ea}{\end{eqnarray}}
\newcommand{\eea}{\end{eqnarray}}
\newcommand{\ord}{\,{\cal O}}

\begin{document}

\title{Doublon-holon pair creation in Mott-Hubbard systems in analogy to QED}

\author{Friedemann Queisser}

\affiliation{Helmholtz-Zentrum Dresden-Rossendorf, 
Bautzner Landstra{\ss}e 400, 01328 Dresden, Germany,}

\affiliation{Institut f\"ur Theoretische Physik, 
Technische Universit\"at Dresden, 01062 Dresden, Germany,}

\author{Konstantin Krutitsky} 

\affiliation{Fakult\"at f\"ur Physik,
Universit\"at Duisburg-Essen, Lotharstra{\ss}e 1, Duisburg 47057, Germany,} 

\author{Patrick Navez} 

\affiliation{Department of Physics, Loughborough University, Loughborough LE11 3TU, United Kingdom} 

\author{Ralf Sch\"utzhold}

\affiliation{Helmholtz-Zentrum Dresden-Rossendorf, 
Bautzner Landstra{\ss}e 400, 01328 Dresden, Germany,}

\affiliation{Institut f\"ur Theoretische Physik, 
Technische Universit\"at Dresden, 01062 Dresden, Germany,}

\date{\today}

\begin{abstract}
Via the hierarchy of correlations, we study doublon-holon pair creation in 
the Mott state of the Fermi-Hubbard model induced by a time-dependent 
electric field.
Special emphasis is placed on the analogy to electron-positron pair creation from 
the vacuum in quantum electrodynamics (QED).
We find that the accuracy of this analogy depends on the spin structure 
of the Mott background. 
For Ising type anti-ferromagnetic order, we derive an effective Dirac equation. 
A Mott state without any spin order, on the other hand, does not explicitly 
display such a quasi-relativistic behavior. 
\end{abstract}

\maketitle

\section{Introduction}

Non-equilibrium dynamics in strongly interacting quantum many-body systems is a
rich and complex field displaying many fascinating phenomena. 
As the {\em drosophila} of strongly interacting quantum many-body systems, 
we consider the Mott insulator phase of the Fermi-Hubbard model \cite{Hub63,Aro22}.
An optical laser can then serve as the external stimulus driving the system 
out of equilibrium -- leading to the creation of doublon-holon pairs, 
see also \cite{Oka03,Eck10,Oka05,Oka10,Avi20,Lig18}.

The intuitive similarity between the upper and lower Hubbard bands 
one the one hand and the Dirac sea and the positive energy continuum in 
quantum electrodynamics (QED) on the other hand suggests analogies 
between doublon-holon pair creation from the Mott state and 
electron-positron pair creation from the vacuum, see also \cite{Sau31,Sch51,Rit85,Niki85,Sch08,Dun09,Pop01,Kohl22,Pop71,Bre70,Kim02,Nar70,Dun05,G05,Nar04}.
In the following, we study this analogy in more detail, with special emphasis 
on the space-time dependence in more than one dimension, such as the 
propagation of doublons and holons. 
More specifically, we strive for an analytic understanding without mapping the 
Fermi-Hubbard Hamiltonian to an effective single site model, see also \cite{Geor92,Eck10}. 

Of course, analogies between electron-positron pair creation in QED and other 
systems at lower energies have already been discussed in previous works. 
Examples include ultra-cold atoms in optical lattices 
\cite{Pin19,Witt11,Cir10,Zhu07,Hou09,Lim08,Boa11,Gold09,Kas16,Quei12,Szp12,Szp11} as well as electrons
in semi-conductors \cite{Smo09,Hri93,Lin18} graphene \cite{
All08,Aka16,Kat12,Novo05,Kat06,Chei06,Vand10,Been08,Fill15,Son09,Ros10,Kao10,Kum23,Schmitt23,Shy09,S84,Dor10,Gav12} and $^3$He \cite{Scho92}.  
However, as we shall see below, there are important differences to the 
Fermi-Hubbard model considered here.
First, the Mott gap arises naturally through the interaction 
(see also \cite{Mott49,Hub63}) 
and does not have to be introduced by hand.  
Second, the particle-hole symmetry between the upper and lower Hubbard band 
-- analogous to the $\cal C$ symmetry in QED -- is also an intrinsic property 
(in contrast to the valence and conduction bands in semi-conductors, for example).
Third, the quantitative analogy to the Dirac equation (in 1+1 dimensions) 
and the resulting quasi-relativistic relativistic behavior does also emerge 
without additional fine-tuning (at least in the case of Ising type spin order, 
see below). 

\section{Extended Fermi-Hubbard Model}

In terms of the fermionic creation and annihilation operators
$\hat c_{\mu s}^\dagger$ and $\hat c_{\nu s}$ 
at the lattice sites $\mu$ and $\nu$ with spin $s\in\{\uparrow,\downarrow\}$ 
and the associated number operators $\hat n_{\mu s}$, 
the extended Fermi-Hubbard Hamiltonian reads ($\hbar=1$) 
\bea
\label{Fermi-Hubbard}
\hat H=-\frac1Z\sum_{\mu\nu s} T_{\mu\nu} \hat c_{\mu s}^\dagger \hat c_{\nu s} 
+U\sum_\mu\hat n_\mu^\uparrow\hat n_\mu^\downarrow
+\sum_{\mu s}V_\mu\hat n_{\mu s}  
\,.\;
\ea
Here the hopping matrix $T_{\mu\nu}$ equals the tunneling strength $T$ 
for nearest neighbors $\mu$ and $\nu$ and is zero otherwise. 
The coordination number $Z$ counts the number of nearest neighbors $\mu$
for a given lattice site $\nu$ and is assumed to be large $Z\gg1$.
In order to describe the Mott insulator, the on-site repulsion $U$ is also 
supposed to be large $U\gg T$.
Finally, the potential $V_\mu(t)$ represents the external electric field, 
e.g., an optical laser. 

\subsection{Hierarchy of Correlations}

To obtain an approximate analytical solution, we consider the reduced density 
matrices of one $\hat\rho_\mu$ and two $\hat\rho_{\mu\nu}$ lattice sites etc.
Next, we split up the correlated parts via 
$\hat\rho_{\mu\nu}^{\rm corr}=\hat\rho_{\mu\nu}-\hat\rho_{\mu}\hat\rho_{\nu}$ 
etc. 
For large $Z\gg1$, we may employ an expansion into powers of $1/Z$ where we find 
that higher-order correlators are successively suppressed \cite{Nav10,Krut14,Queiss19,Queiss14,Nav16}. 
More precisely, the two-point correlator scales as 
$\hat\rho_{\mu\nu}^{\rm corr}=\ord(1/Z)$,
while the three-point correlation is suppressed as 
$\hat\rho_{\mu\nu\lambda}^{\rm corr}=\ord(1/Z^2)$ etc.   

Via this expansion into powers of $1/Z$, we may find approximate 
solutions of the evolution equations 
\bea
\label{evolution}
i\partial_t \hat\rho_\mu 
&=& 
F_1(\hat\rho_\mu,\hat\rho_{\mu\nu}^{\rm corr})
\,,\nn
i\partial_t \hat\rho_{\mu\nu}^{\rm corr} 
&=& 
F_2(\hat\rho_\mu,\hat\rho_{\mu\nu}^{\rm corr},\hat\rho_{\mu\nu\lambda}^{\rm corr})
\,.
\ea
Using $\hat\rho_{\mu\nu}^{\rm corr}=\ord(1/Z)$, 
the first evolution equation can be approximated by 
$i\partial_t \hat\rho_\mu = F_1(\hat\rho_\mu,0)+\ord(1/Z)$.  
Its zeroth-order solution $\hat\rho_\mu^0$ yields the mean-field background,
which will be specified below. 

Next, the suppression $\hat\rho_{\mu\nu\lambda}^{\rm corr}=\ord(1/Z^2)$ 
allows us to approximate the second equation~\eqref{evolution} to leading order 
in $1/Z$ via $i\partial_t \hat\rho_{\mu\nu}^{\rm corr}\approx 
F_2(\hat\rho_\mu^0,\hat\rho_{\mu\nu}^{\rm corr},0)$. 
In order to solve this leading-order equality, it is convenient to split to 
fermionc creation and annihilation operators in particle $I=1$ 
and hole $I=0$ contributions via 
\bea
\hat c_{\mu s I}=\hat c_{\mu s}\hat n_{\mu\bar s}^I=
\left\{
\begin{array}{ccc}
 \hat c_{\mu s}(1-\hat n_{\mu\bar s}) & {\rm for} & I=0 
 \\ 
 \hat c_{\mu s}\hat n_{\mu\bar s} & {\rm for} & I=1
\end{array}
\right.
\,,
\ea
where $\bar s$ denotes the spin opposite to $s$. 
In terms of these particle and hole operators, the correlations 
(for $\mu\neq\nu$) are determined by 
\bea
\label{correlations}
i\partial_t
\langle\hat c^\dagger_{\mu s I}\hat c_{\nu s J}\rangle^{\rm corr}
=
\frac1Z\sum_{\lambda L} T_{\mu\lambda}
\langle\hat n_{\mu\bar s}^I\rangle^0
\langle\hat c^\dagger_{\lambda s L}\hat c_{\nu s J}\rangle^{\rm corr}
\nn
-
\frac1Z\sum_{\lambda L} T_{\nu\lambda}
\langle\hat n_{\nu\bar s}^J\rangle^0
\langle\hat c^\dagger_{\mu s I}\hat c_{\lambda s L}\rangle^{\rm corr}
\nn
+
\left(U_\nu^J-U_\mu^I+V_\nu-V_\mu\right) 
\langle\hat c^\dagger_{\mu s I}\hat c_{\nu s J}\rangle^{\rm corr}
\nn
+\frac{T_{\mu\nu}}{Z}
\left(
\langle\hat n_{\mu\bar s}^I\rangle^0
\langle\hat n_{\nu s}\hat n_{\nu\bar s}^J\rangle^0
-
\langle\hat n_{\nu\bar s}^J\rangle^0
\langle\hat n_{\mu s}\hat n_{\mu\bar s}^I\rangle^0
\right) 
\,,
\ea
where $\langle\hat X_\mu\rangle^0={\rm Tr}\{\hat X_\mu\hat\rho_\mu^0\}$
denote expectation values in the mean-field background.


The evolution equations~\eqref{correlations} for the correlators can be simplified 
by factorizing them via the following effective linear equations for the particle 
and hole operators
\bea
\label{factorization}
\left(i\partial_t-U^I_\mu-V_\mu\right)\hat c_{\mu s I}
=
-\frac1Z\sum_{\nu J} 
T_{\mu\nu} \langle\hat n_{\mu\bar s}^I\rangle^0 \hat c_{\nu s J}
\,.
\ea
Of course, the hierarchy of correlations is not the only way to derive such 
effective evolution equations, similar results can be obtained by other 
approximation schemes, e.g., \cite{Fis08}. 

\section{Ising type spin order}

In order to analyze the effective equations~\eqref{factorization}, we have to 
specify the mean-field background $\hat\rho_\mu^0$. 
The Mott insulator state corresponds to having one particle per lattice site,
which leaves to determine the remaining spin degrees of freedom. 
As our first example, we consider anti-ferromagnetic spin order of the Ising type \cite{Hir89}. 
To this end, we assume a bi-partite lattice which can be spit into two sub-lattices 
$\cal A$ and $\cal B$ where all neighbors $\nu$ of a lattice site $\mu\in\cal A$
belong to $\cal B$ and vice versa. 
Then, the zeroth-order mean-field background reads 
\bea
\label{mean-field-Ising}
\hat\rho_\mu^0
=
\left\{ 
\begin{array}{ccc}
\ket{\uparrow}_\mu\!\bra{\uparrow} & {\rm for} & \mu\in\cal A
\\
\ket{\downarrow}_\mu\!\bra{\downarrow} & {\rm for} & \mu\in\cal B
\end{array}
\right. 
\,.
\ea
This state
minimizes the Ising type anti-ferromagnetic interaction 
$\hat S^z_\mu\hat S^z_\nu$. 
Note that the Fermi-Hubbard Hamiltonian~\eqref{Fermi-Hubbard} does indeed 
generate an effective anti-ferromagnetic interaction via second-order hopping 
processes, but it would correspond to a Heisenberg type anti-ferromagnetic 
interaction $\hat{\f{S}}_\mu\cdot\hat{\f{S}}_\nu$ \cite{Cha78}. 
Although the state~\eqref{mean-field-Ising} does not describe the exact minimum 
of this interaction $\hat{\f{S}}_\mu\cdot\hat{\f{S}}_\nu$, it can be regarded as 
an approximation or a simplified toy model for such an anti-ferromagnet.
Alternatively, one could imagine additional spin interactions between the 
electrons (stemming from the full microscopic description) which are not 
contained in the tight-binding model~\eqref{Fermi-Hubbard} and stabilize 
the state~\eqref{mean-field-Ising}. 

\subsection{Effective Dirac Equation}

This background~\eqref{mean-field-Ising} supports 
hole excitations $I=0$ of spin $\uparrow$ 
and particle excitations $I=1$ of spin $\downarrow$ 
for the sub-lattice $\cal A$, and vice versa for the sub-lattice $\cal B$. 
For the other terms, such as $\hat c_{\mu\in{\cal A},s=\uparrow,I=1}$, 
the right-hand side of Eq.~\eqref{factorization} vanishes and thus they 
become trivial and are omitted in the following.

As a result, particle excitations in the sub-lattice $\cal A$ are 
tunnel coupled to hole excitations of the same spin in the sub-lattice $\cal B$
and vice versa. 
Since the spin components $s=\uparrow$ and $s=\downarrow$ evolve independent 
of each other, we drop the spin index in the following. 
Introducing the effective spinor in analogy to the Dirac equation 
\bea
\hat\psi_\mu
=
\left(
\begin{array}{c}
\hat c_{\mu  I=1} 
\\
\hat c_{\mu  I=0} 
\end{array}
\right) 
\,,
\ea
the evolution equation can be cast into the form 
\bea
\label{pre-Dirac}
i\partial_t\hat\psi_\mu
=
\left(
\begin{array}{cc} 
V_\mu+U & 0 
\\
0 & V_\mu 
\end{array}
\right)
\cdot\hat\psi_\mu
-\frac1Z\sum_{\nu}T_{\mu\nu}
\sigma_x\cdot\hat\psi_\nu
\,,
\ea
where $\sigma_x$ is the Pauli spin matrix. 
This form is already reminiscent of the Dirac equation in 1+1 dimensions.  
To make the analogy more explicit, we first apply a simple phase 
transformation $\hat\psi_\mu\to\hat\psi_\mu\exp\{itU/2\}$ after which 
the $U$ term reads $\sigma_zU/2$. 

Since the wavelength of an optical laser is typically much longer than all 
other relevant length scales, we may approximate it 
(in the non-relativistic regime) by a purely time-dependent electric field 
$\f{E}(t)$ such that the potential reads $V_\mu(t)=q\f{r}_\mu\cdot\f{E}(t)$
where $\f{r}_\mu$ is the position vector of the lattice site $\mu$. 
Then we may use the Peierls transformation 
$\hat\psi_\mu\to\hat\psi_\mu\exp\{i\varphi_\mu(t)\}$ with 
$\dot\varphi_\mu=V_\mu$ to shift the potential $V_\mu$ into 
time-dependent phases of the hopping matrix 
$T_{\mu\nu}\to T_{\mu\nu}e^{i\varphi_\mu(t)-i\varphi_\nu(t)}=T_{\mu\nu}(t)$.
Next, a spatial Fourier transformation simplifies Eq.~\eqref{pre-Dirac} to 
\bea
\label{Dirac-k}
i\partial_t\hat\psi_{\fk{k}}
=
\left(
\frac{U}{2}\,\sigma_z
-
T_{\fk{k}}(t)\sigma_x
\right)\cdot\hat\psi_{\fk{k}}
\,,
\ea
where $T_{\fk{k}}(t)$ denotes the Fourier transform of the hopping matrix 
including the time-dependent phases, which yields the usual minimal coupling 
form $T_{\fk{k}}(t)=T_{\fk{k}-q\fk{A}(t)}$. 
Note that the Peierls transformation is closely related to the gauge 
transformation $A_\mu\to A_\mu+\partial_\mu\chi$ in electrodynamics. 
Using this gauge freedom, one can represent an electric field $\f{E}(t)$
via the scalar potential $\phi$ as $\partial_t+iq\phi$ in analogy to the $V_\mu$  
or via the vector potential $\f{A}$ as $\nabla-iq\f{A}$ in analogy to the
$T_{\fk{k}}(t)=T_{\fk{k}-q\fk{A}(t)}$. 

In the absence of the electric field, the dispersion relation following from 
Eq.~\eqref{Dirac-k} reads 
\bea
\label{dispersion-relativistic}
\omega_{\fk{k}}=\pm\sqrt{\frac{U^2}{4}+T_{\fk{k}}^2}
\,.
\ea
The positive and negative frequency solutions correspond to the upper and lower 
Hubbard bands, which are separated by the Mott gap. 
Unless the electric field is too strong or too fast, one expects the main 
contributions to doublon-holon pair creation near the minimum gap, i.e., 
the minimum of $T_{\fk{k}}^2$, typically at $T_{\fk{k}}=0$.
Then, a Taylor expansion $\f{k}=\f{k}_0+\delta\f{k}$
around a zero $\f{k}_0$ of $T_{\fk{k}}$ yields 
\bea
\label{Dirac-approx}
i\partial_t\hat\psi_{\fk{k}}
\approx 
\left(
\frac{U}{2}\,\sigma_z
-
\f{c}_{\rm eff}\cdot[\delta\f{k}-q\f{A}(t)] 
\sigma_x
\right)\cdot\hat\psi_{\fk{k}}
\,,
\ea
where $\f{c}_{\rm eff}=\nabla_{\fk{k}}T_{\fk{k}}|_{\fk{k}_0}$ 
denotes the effective propagation velocity, 
in analogy to the speed of light.  
Note that the validity of this approximation does not only require 
$\delta\f{k}$ to be small, it also assumes that $q\f{A}(t)$ does not become 
too large, e.g., that we are far away from the regime of Bloch oscillations \cite{Eck11,Niu96}.  

On the other hand, even if $q\f{A}(t)$ varies over a larger range, 
the above approximation~\eqref{Dirac-approx} could still provide a reasonably 
good description for strong-field doublon-holon pair creation. 
This process can be understood as Landau-Zener tunneling occurring when an 
avoided level crossing is traversed with a finite speed 
(set by $\f{E}=-\dot{\f{A}}$). 
Since this tunneling process mainly occurs in the vicinity of the minimum gap, 
i.e., where $T_{\fk{k}}(t)=T_{\fk{k}-q\fk{A}(t)}=0$, 
it is sufficient to consider the region around $\f{k}-q\f{A}=\f{k}_0$. 

\subsection{Analogy to QED}

Up to simple phase factors $\exp\{i\f{k}_0\cdot\f{r}_\mu\}$, 
Eq.~\eqref{Dirac-approx} displays a qualitative analogy to the Dirac equation 
in 1+1 dimensions, where $c_{\rm eff}$ plays the role of the speed of light 
while $U/2$ corresponds to the mass $m_{\rm eff}c_{\rm eff}^2=U/2$. 
As a result, we may now apply many of the results known from 
quantum electrodynamics (QED) \cite{
Dirac1,Dirac2,Klein29,Sau32,Hei36,Zen32,Zen34,Lan32,Bun70,Niki67,Kel65}. 
For example, the effective spinor $\hat\psi_{\fk{k}}$ can be expanded into  
particle and hole contributions (first in the absence of an electric field) 
\bea
\hat\psi_{\fk{k}}
=
u_{\fk{k}}\hat d_{\fk{k}}+v_{\fk{k}}\hat h_{\fk{k}}^\dagger 
\,,
\ea
where the quasi-particles are usually referred to as doublons $\hat d_{\fk{k}}$
(upper Hubbard band) and holons $\hat h_{\fk{k}}^\dagger$
(lower Hubbard band). 
The Mott state $\ket{\rm Mott}$ is then determined by 
$\hat d_{\fk{k}}\ket{\rm Mott}=\hat h_{\fk{k}}\ket{\rm Mott}=0$.

In the presence of an electric field, these operators can mix -- 
as described by the Bogoliubov transformation 
\bea
\hat d_{\fk{k}}^{\rm out}
=
\alpha_{\fk{k}}\hat d_{\fk{k}}^{\rm in}
+
\beta_{\fk{k}}\left(\hat h_{\fk{k}}^{\rm in}\right)^\dagger 
\,,
\ea
where the structure of the Dirac equation~\eqref{Dirac-approx} implies the 
normalization $|\alpha_{\fk{k}}|^2+|\beta_{\fk{k}}|^2=1$. 
Starting in the Mott state 
$\hat d_{\fk{k}}^{\rm in}\ket{\rm Mott}=\hat h_{\fk{k}}^{\rm in}\ket{\rm Mott}=0$,
the $\beta_{\fk{k}}$ coefficient yields the amplitude for doublon-holon pair 
creation 
$\hat d_{\fk{k}}^{\rm out}\ket{\rm Mott}\propto\beta_{\fk{k}}$. 
In analogy to QED, we may now discuss different regimes. 
For weak electric fields $\f{E}$ oscillating near resonance $\omega\approx U$,
we find the usual lowest-order perturbative scaling 
$\beta_{\fk{k}}\sim|q\f{c}_{\rm eff}\cdot\f{E}|$ \cite{Bre34}. 
Higher orders of perturbation theory then lead us in the multi-photon regime, 
for example $n\omega\approx U$ with 
$\beta_{\fk{k}}\sim|q\f{c}_{\rm eff}\cdot\f{E}|^n$. 

Note that a completely different kind of resonances such as $\omega\approx2U$ 
can occur if we take higher-order correlations into account \cite{Quei19,Quei19b}, but these 
are beyond our effective description~\eqref{Dirac-approx}.

If the electric field becomes stronger and slower, we enter the non-perturbative
regime of the Sauter-Schwinger effect where the pair-creation amplitude displays 
an exponential scaling \cite{Sau31,Sch51}
\bea
\beta_{\fk{k}}\sim\exp\left\{-\frac{\pi U^2}{8q|\f{c}_{\rm eff}\cdot\f{E}|}\right\}
\,.
\ea
The quantitative analogy to the Dirac equation even allows us to directly 
transfer further results from QED, for example the dynamically assisted 
Sauter-Schwinger effect, where pair creation by a strong and slowly varying 
electric field is enhanced by adding a weaker and faster varying field, see,
e.g., \cite{Sch08}.

\section{Unordered spin state}

Let us compare our findings above to the case of a mean-field 
background without any spin order
\bea
\label{mean-field-unordered}
\hat\rho_\mu^0
=
\frac{\ket{\uparrow}_\mu\!\bra{\uparrow}+\ket{\downarrow}_\mu\!\bra{\downarrow}}{2}
\,, 
\ea
which could arise for a finite temperature which is too small to excite 
doublon-holon pairs but large enough to destroy the spin order. 
Another option could be a weak magnetic disorder potential and/or spin 
frustration. 

In this case, we do not have to distinguish the two sub-lattices $\cal A$ 
and $\cal B$ and all lattice sites can support particle and hole excitations.
Since all expectation values $\langle\hat n_{\mu s}^I\rangle^0$ yield $1/2$ 
(instead of zero or unity), the analog of Eq.~\eqref{Dirac-k} now reads 
(after the same phase transformation)
\bea
\label{matrix-unordered}
i\partial_t
\hat\psi_{\fk{k}}
=
\frac12
\left(
\begin{array}{cc}
U-T_{\bf k} & -T_{\bf k}
\\
-T_{\bf k} & -U-T_{\bf k}
\end{array}
\right)
\cdot
\hat\psi_{\fk{k}}
%
\,.
\ea
The eigenvalues of the above $2\times2$-matrix yield the 
quasi-particle frequencies \cite{Hub63,Herr97}
\bea
\label{quasi-particle-energies}
\omega^\pm_\mathbf{k}
=
\frac12\left(-T_\mathbf{k}\pm\sqrt{T_\mathbf{k}^2+U^2}\right)
\,.
\ea
In view of the additional term $-T_\mathbf{k}$ in front of the square root, 
this dispersion relation does not display the same quasi-relativistic form 
as in Eq.~\eqref{dispersion-relativistic}.
Putting it another way, we find that Eq.~\eqref{matrix-unordered}
is not formally equivalent to the Dirac equation (in 1+1 dimensions). 

As a consequence, the propagation of quasi-particles in the two mean-field 
backgrounds~\eqref{mean-field-Ising} and \eqref{matrix-unordered} is quite 
different. 
For the Ising type order~\eqref{mean-field-Ising}, the coherent propagation 
of a doublon or holon (without changing the background structure) 
requires second-order hopping processes. 
Hence $\omega_{\fk{k}}$ in Eq.~\eqref{dispersion-relativistic} 
is a quadratic function of $T_{\fk{k}}$.
For the unordered background~\eqref{mean-field-unordered}, on the other hand, 
doublons and holons can propagate coherently via first-order hopping processes. 
This is reflected in the linear contribution $-T_\mathbf{k}$ in 
Eq.~\eqref{quasi-particle-energies}.

However, if we do not consider quasi-particle propagation but focus on the 
probability for creating a doublon-holon pair in a given mode $\f{k}$,
we may again derive a close analogy to QED. 
To this end, we apply yet another $\f{k}$-dependent phase transformation 
$\hat\psi_{\fk{k}}\to e^{i\vartheta_{\fk{k}}(t)}\hat\psi_{\fk{k}}$ with 
$\dot\vartheta_{\fk{k}}(t)=T_{\fk{k}}(t)/2$.
Note that $T_{\fk{k}}(t)$ contains the vector potential $\f{A}(t)$, i.e., 
the time integral of the electric field $\f{E}(t)$.
Thus, the phase $\vartheta_{\fk{k}}(t)$ involves yet another time integral,
which makes it a even more non-local function of time.
After this phase transformation, Eq.~\eqref{matrix-unordered} becomes again 
formally equivalent to the Dirac equation in 1+1 dimensions, but now with 
the effective speed of light being reduced by a factor of two.

\section{Conclusions}

We study doublon-holon pair creation from the Mott insulator state of the 
Fermi-Hubbard model induced by an external electric field $\f{E}(t)$ which 
could represent an optical laser, for example.
We find that the creation and propagation dynamics of the doublons and holons 
depends on the spin structure of the mean-field background.  
For Ising type anti-ferromagnetic order, we observe a quantitative analogy 
to QED.
More specifically, in the vicinity of the minimum gap 
(i.e., the most relevant region for pair creation), 
the doublons and holons are described by an effective
Dirac equation in 1+1 dimensions in the presence of an electric field 
$\f{E}(t)$.

As a consequence of this quantitative analogy, we may employ the machinery of QED 
and apply many results regarding electron-positron pair creation to our set-up. 
For example, in the perturbative (single- or multi-photon) regime with the  
threshold conditions $\omega\geq nU$ for the $n$-th order, the doublon-holon 
pair creation amplitude $\beta_{\fk{k}}$ yields the perturbative scaling 
$\beta_{\fk{k}}\sim|q\f{c}_{\rm eff}\cdot\f{E}|^n$. 

For stronger and slower electric fields, we enter the non-perturbative (tunneling) 
regime in analogy to the Sauter-Schwinger effect in QED and thus recover 
the exponential dependence already discussed earlier regarding the dielectric 
breakdown of Mott insulator, see, e.g., \cite{Eck10}. 
Note that the quantitative analogy established above (see also Table~\ref{table1})
unambiguously determines the 
pair-creation exponent and pre-factor without and free fitting parameters. 
%

If we consider the annihilation of doublon-holon pairs instead of their 
creation, this analogy applies to the stimulated annihilation within an 
external field, but not to the spontaneous annihilation of an electron-positron
pair by emitting a pair of photons, for example. 
In order to model this process, one has to include a mechanism for dissipating
the energy, e.g., by coupling the Fermi-Hubbard model to an environment, see 
also \cite{Quei19c}.

For a mean-field background without any spin order, on the other hand, 
the creation and propagation of doublons and holons does not display such 
a quasi-relativistic behavior. 
The dispersion relation is different and the evolution equation deviates from 
the Dirac equation. 
Still, for a purely time-dependent electric field $\f{E}(t)$ considered here,
the doublon-holon pair creation amplitude $\beta_{\fk{k}}$ for a given mode 
$\f{k}$ can again be related to QED.
After a $\f{k}$-dependent phase transformation (which is non-local in time), 
the amplitude $\beta_{\fk{k}}$ is given by the same expression, just with 
the effective speed of light $c_{\rm eff}$ being reduced by a factor of two. 

In view of the $\f{k}$-dependence of the phase transformation, this mapping 
does only work for purely time-dependent electric field $\f{E}(t)$.
For space-time dependent electric fields $\f{E}(t,\f{r})$, the deviation of 
the dispersion relation and the resulting difference in propagation become 
important -- which will be the subject of further studies.  

\begin{table}[ht] 
\begin{tabular}{|c|c|}
\hline 
Mott insulator & QED vacuum 
\\
upper Hubbard band & positive energy continuum 
\\ 
lower Hubbard band & Dirac sea 
\\
doublons \& holons & electrons \& positrons 
\\
Mott gap $U=2m_{\rm eff}c_{\rm eff}^2$ & electron mass
\\
velocity $\f{c}_{\rm eff}=\nabla_{\fk{k}}T_{\fk{k}}|_{\fk{k}_0}$ 
& speed of light $c$ 
\\
Landau-Zener tunneling & Sauter-Schwinger effect 
\\
Peierls transformation & gauge transformation
\\
\hline 
\end{tabular}
\label{table1}
\caption{Sketch of the analogy. 
The electric field $\f{E}(t)$ and elementary charge $q$ play the same role 
in both cases.}
\end{table}

\acknowledgments

%
Funded by the Deutsche Forschungsgemeinschaft (DFG, German Research Foundation) 
-- Project-ID 278162697-- SFB 1242. 
PN thanks support from the EU project
SUPERGALAX (Grant agreement ID: 863313).

\end{document}